# A Comprehensive Model of Snow Crystal Faceting

Kenneth Libbrecht[1] and James Walkling[2]

[1]Department of Physics, California Institute of Technology
Pasadena, California 91125, kgl@caltech.edu

[2]Department of Physics, University of Cambridge
Cambridge, England CB3 0HE

**Abstract.** Crystal faceting can emerge via two broad physical mechanisms: anisotropic attachment kinetics on growing crystals and anisotropic surface energies on near-equilibrium crystals. For the case of the ice/vapor system, anisotropic attachment kinetics is the dominant faceting mechanism, while the possible occurrence of equilibrium faceting has been debated for many decades. In this investigation we examine ice/vapor faceting at low supersaturations over the temperature range $-15C < T < 0C$, where evidence of a roughening transition has been previously reported. Our findings indicate that a comprehensive attachment kinetics model can explain all the experimental data to date, while assuming an essentially isotropic surface energy (which is supported by other considerations). Specifically, our kinetic model naturally explains the observed disappearance of prism faceting on slowly growing ice crystals in vacuum at $T > -2C$, thus suggesting that snow crystal faceting is caused by anisotropic attachment kinetics even at extremely slow growth rates.

## ❋ Introduction

Natural crystal facets are observed on many mineral crystals, with ice and quartz being two of the most common examples. In most mineral systems, faceted surfaces emerge during crystal growth involving anisotropic attachment kinetics, which is an intrinsically non-equilibrium process. Specifically, facet surfaces (having low Miller indices) accumulate material more slowly than other surfaces, with growth often being limited by terrace nucleation on the molecularly smooth facets. In this situation, the slowest-growing surfaces typically define the overall growth morphology, yielding faceted growth forms. For the specific case of ice growing from water vapor (snow crystals), hexagonal prisms are the simplest and most common fully faceted morphology, although pyramidal facets also sometimes appear at low temperatures [2006Tap, 2021Lib].

In generally rarer circumstances, crystal facets can also appear in the absence of growth, where the Equilibrium Crystal Shape (ECS) is determined by minimizing the total surface energy of an isolated test crystal. If the surface energies on faceted surfaces are substantially lower than on non-faceted (rough) surfaces, then the ECS will be faceted [1980Hey, 1987Hey]. There is much discussion of faceted ECSs in the scientific literature, and it is often thought that the ECS for the ice/vapor system is a faceted hexagonal prism [1997Pru]. However, the available evidence to date suggests that the ice/vapor surface-energy anisotropy is quite small at temperatures above -15 C, so the snow-crystal ECS is very nearly spherical in this temperature range [2012Lib2].

Faceting observed in snow crystals typically arises from nucleation barriers that greatly suppress the growth of faceted basal and prism surfaces [1982Kur, 1984Kur, 1984Kur1, 1987Kob, 1998Nel, 2021Lib]. As shown below, the terrace nucleation mechanism yields exceedingly slow growth rates at low supersaturations, especially at low temperatures, often yielding highly faceted growth forms. When small ice crystals are grown from water vapor in near-vacuum conditions, the growth forms are typically simple hexagonal prisms.

The experimental situation becomes a bit confusing at temperatures above -2 C, however, as we describe in detail below. Basal faceting remains pronounced at all temperatures, but prism faceting is present in some circumstances while remaining absent in others. For example, we have observed strong prism faceting in air at temperatures as high as -0.2 C [2021Lib2], while prism faceting at -2 C is sometimes (but not always) substantially reduced for crystals grown in near-vacuum conditions. Similar observations by other researchers have been interpreted as evidence for changes in the ECS with temperature [1985Col], and perhaps a roughening transition on prism surfaces [1991Elb]. Overall, the experimental observations have not painted a clear picture of ice faceting behavior, and the ice/vapor ECS remains a topic of scientific debate.

Our overarching goal in this paper is to develop a comprehensive model of faceting in the ice/vapor system, focusing especially on simple faceted prisms that appear at low growth rates. From the outset we assume that the surface-energy anisotropy negligibly small, so the ice/vapor ECS is essentially spherical. The available evidence suggests that the real ECS likely exhibits only minute facets on an otherwise spherical form [2012Lib2], supporting our spherical approximation. We also assume a terrace-nucleation model to describe growth of the basal and prism facets using model parameters determined from experimental ice-growth measurements [2013Lib, 2021Lib].

With these model assumptions, we find that we can explain essentially all the available experimental observations to a reasonable degree, with the caveat that there remain substantial uncertainties in both the experiments observations and our model calculations. Our model shows that faceting from anisotropic attachment kinetics is important in all but the most extreme conditions, and that an anisotropic surface energy is not a necessary requirement to explain the existing data.

Importantly, our model establishes a theoretical framework for further investigations of snow crystal faceting, and for further consideration of the ice/vapor ECS and how it could be definitively observed. The model therefore makes important progress in the continuing exploration of crystal growth dynamics in the ice/vapor system, particularly under physical conditions approaching the triple point.

## ❄ A Basic Analytic Model of Snow Crystal faceting

During snow crystal formation, a variety of physical processes influence the growth dynamics, including attachment kinetics, particle and heat diffusion, and surface energy effects [2021Lib]. The formation of snow crystals in air is mainly governed by the interplay of particle diffusion and attachment kinetics, typically yielding complex morphologies that are both branched and faceted. In this paper, we focus our attention on slow growth that yields simple faceted prisms, especially in low-pressure experiments, where particle diffusion plays a relatively small role. We begin our model development by defining a suitable parameterization of the growth dynamics and attachment kinetics.

The basic tenets of molecular attachment kinetics have been generally understood for about a century [1882Her, 1915Knu, 1990Yok] and are explained in numerous textbooks describing the physics of crystal growth

[1990Sai, 1999Pim, 2004Mar]. For the ice/vapor system we write the Hertz-Knudsen relation [2021Lib]

$$v_n = \alpha v_{kin} \sigma_{surf} \quad (1)$$

for the growth of a flat surface, where $v_n$ is the crystal growth velocity perpendicular to the growing surface, $\alpha$ is a dimensionless *attachment coefficient*, $\sigma_{surf} = (c_{surf} - c_{sat})/c_{sat}$ is the dimensionless water vapor supersaturation at the surface, $c_{surf}$ is the water-vapor number density just above the surface, $c_{sat} = c_{sat}(T)$ is the saturated number density of a surface in equilibrium at temperature $T$, and

$$v_{kin} = \frac{c_{sat}}{c_{ice}} \sqrt{\frac{kT}{2\pi m_{mol}}} \quad (2)$$

is the *kinetic velocity*, in which $m_{mol}$ is the mass of a water molecule, $c_{ice} = \rho_{ice}/m_{mol}$ is the number density of ice, and $\rho_{ice}$ is the mass density of ice.

For a non-faceted (a.k.a. rough) ice surface, measurements indicate $\alpha_{rough} \approx 1$ under most conditions [2021Lib], while $\alpha_{facet} \ll 1$ when $\sigma_{surf}$ is low, yielding strong basal and prism faceting over a broad range of environmental conditions.

## Attachment Kinetics

Measurements of $\alpha_{facet}$ have shown that the attachment kinetics are primarily limited by terrace nucleation under typical growth conditions [2013Lib], prompting us to write $\alpha_{facet}$ as [1996Sai, 2021Lib]

$$\alpha_{facet}(\sigma_{surf}) = Ae^{-\sigma_0/\sigma_{surf}} \quad (3)$$

where $A(T)$ and $\sigma_0(T)$ are dimensionless parameters that are generally different for the basal and prism facets, with

$$\sigma_0(T) = \frac{S\beta^2 a^2}{k^2 T^2} \quad (4)$$

where $a$ is the molecular size, $k$ is the Boltzmann factor, $T$ is the surface temperature, and $\beta$ is the step energy of a terrace edge. This expression includes a dimensionless geometrical factor $S \approx 1$ to absorb several small theoretical factors [2021Lib].

The validity of this functional form for terrace nucleation in the ice/vapor system has been verified by experiments over a broad range of temperatures and supersaturations [2013Lib], and the current state-of-the-art for measurements of $A(T)$ and $\sigma_0(T)$ on both the basal and prism facets is presented in considerable detail in [2021Lib] and references therein.

## Surface energy effects

Snow crystal growth rates are influenced by the ice/vapor interfacial energy mainly via the Gibbs-Thomson effect, which describes how the equilibrium vapor pressure above a curved surface is higher than that above a flat surface. Again this is a well-known result in statistical mechanics and crystal-growth theory, yielding a modified Hertz-Knudsen relation [1996Sai, 2021Lib]

$$v_n = \alpha v_{kin}(\sigma_{surf} - d_{sv}\kappa) \quad (5)$$

where

$$d_{sv} = \frac{\gamma_{sv}}{c_{ice}kT} \approx 1\ nm \quad (6)$$

is the Gibbs-Thomson length, $\gamma_{sv}$ is the surface energy of the solid/vapor interface, and $\kappa$ is the local surface curvature. For a spherical surface we have $\kappa = 2/R$ where $R$ is the radius of the sphere. Note that $\kappa = 0$ only for flat surfaces of infinite size, and we must assume $\kappa > 0$ for any facets on finite-size test crystals. For an approximately isometric hexagonal prism (a common experimental case) with an overall effective radius approximately equal to $R$, we take $\kappa \approx 2/R$ for the facet surfaces.

We assume that $\gamma_{sv}$ has a constant value independent of surface orientation in this paper, implying a spherical ECS for the ice/vapor system. The evidence suggests that this is an excellent approximation at temperatures above -15 C, although there have been no definitive measurements of the ice ECS at any temperature to date (in our opinion). Questions relating to the ice ECS and the anisotropy of the ice/vapor surface energy remain a topic of scientific debate.

The nucleation dynamics on faceted surfaces of finite size will be affected by the Gibbs-Thomson effect, as there can be no nucleation if the effective radius of a facet is smaller than the critical terrace radius in nucleation theory. The theoretical questions that arise in such circumstances are beyond the scope of this paper, but our investigation suggests that this effect is rather small, being significant only for exceptionally small crystals held at very low surface supersaturations. Nevertheless, we approximate the resulting changes by modifying the attachment kinetics on faceted surfaces to be

$$\alpha_{facet}(\sigma_{surf}) = Ae^{-\sigma_0/(\sigma_{surf}-d_{sv}\kappa_{facet})} \quad (7)$$

## Particle diffusion

Snow crystal growth is often strongly limited by the slow diffusion of water vapor molecules through the surrounding medium (usually air), and a full 3D solution to the problem of diffusion-limited growth remains a challenging computational task [2021Lib]. However, the spherical case has a simple analytic solution that can be useful for approximating the supersaturation field around a nearly isometric hexagonal prism. In this one-dimensional diffusion problem, the supersaturation at all points can be written [2021Lib]

$$\sigma(r) = \sigma_\infty - \frac{R}{r}(\sigma_\infty - \sigma_{surf}) \quad (8)$$

where

$$\sigma_{surf} = \frac{\alpha_{diff}}{\alpha + \alpha_{diff}}\sigma_\infty \quad (9)$$

with

$$\alpha_{diff} = \left(\frac{c_{sat}}{c_{ice}}\frac{D}{v_{kin}}\right)\frac{1}{R}$$
$$= \frac{X_0}{R} \quad (10)$$

and $D$ is the diffusion constant, giving the crystal growth velocity

$$v_n = \left(\frac{\alpha\alpha_{diff}}{\alpha + \alpha_{diff}}\right)v_{kin}\sigma_\infty \quad (11)$$

Rearranging these expressions gives

$$\sigma(r) = \sigma_\infty - \frac{1}{4\pi\rho_{ice}X_0 v_{kin}}\frac{dM}{dt}\frac{1}{r} \quad (12)$$

where $dM/dt$ is the rate of change of the mass of the crystal.

At very slow growth rates Equation (12) becomes $\sigma(r) \to \sigma_\infty$ at all $r$, as one would expect, and this expression can provide a reasonable estimate of the supersaturation field around a growing prism provided $\alpha \ll \alpha_{diff}$. The approximation becomes less useful when this inequality becomes less valid, and it becomes essentially useless when $\alpha_{diff} < \alpha$.

Looking at our central question of faceting in this paper, we see that particle diffusion tends to enhance the formation of faceted forms with sharp corners and edges via the Mullins-Sekerka instability [1964Mul, 2021Lib]. With a hexagonal prism, for example, the hexagonal tips stick out farther into the supersaturated surroundings, so the increased supersaturation associated with particle diffusion tends to promote the tip growth. At high growth rates and with large crystals, this effect leads to branching and complex growth morphologies. At low growth rates and with small crystals, particle diffusion may do little more than slightly encourage the formation of faceted forms.

## Heat diffusion

In a low-pressure experimental environment, the particle diffusion constant becomes quite large, and under these conditions the effects of particle diffusion often become negligible. In these same circumstances, however, thermal diffusion often becomes an important factor limiting crystal growth. With a particle resting on a substrate, for example, latent heat released during growth often dissipates by being conducted through the ice to the supporting substrate. The resulting heat flow produces a temperature gradient within the crystal, with the top surface being warmer than the part of the crystal contacting the substrate. The morphological effects from this heating are often seen when ice crystals are grown in a near-vacuum environment [1972Lam].

As with particle diffusion, the full 3D heat diffusion problem can be quite challenging to solve. Fortunately, 1D analytic solutions again provide useful insights into the overall scale of the problem, and they can be used to reasonably estimate of how latent heating affects crystal growth rates in experiments. For the case of a uniform sheet of ice growing on a substrate [2021Lib], the growth rate can be written

$$v_n = \left(\frac{\alpha \alpha_{therm}}{\alpha + \alpha_{the}}\right) v_{kin} \sigma_\infty \quad (13)$$

with

$$\alpha_{therm} = \frac{\kappa_{ice}}{\eta L_{sv} \rho_{ice} v_{kin}} \frac{G}{H} \quad (14)$$

where $H$ is the thickness of the sheet, $G = 1$, and

$$\eta = \frac{1}{c_{sat}} \frac{dc_{sat}}{dT} \quad (15)$$

For the case of a small hexagonal prism growing on a substrate, this same expression can approximate the growth when $G$ is replaced by a geometrical constant of order unity. Thus we can use this simple analytic expression to approximately model the full effects of latent heating in practical experimental situations.

In contrast to particle diffusion, thermal diffusion tends to hinder the formation of faceted forms with sharp corners and edges. For the case of a simple prism growing on a substrate, the combination of perpendicular and lateral growth yields the highest crystal temperature increases at the tips and edges farthest from the substrate, so these sharp structures will tend to round from latent heating effects [1972Lam].

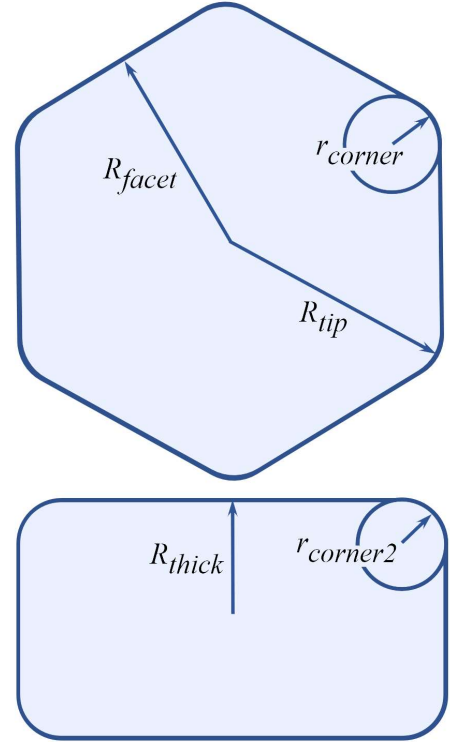

Figure 1. The geometry of a simple hexagonal prism with edges and corners rounded by the Gibbs-Thomson effect.

## ❄ Slow growth of simple ice prisms in vacuum

With the model underpinnings described above, let us now consider the slow growth of a simple ice prism defined by the geometrical parameters in Figure 1. If the crystal rests on an inert heat-conducting substrate in a low-pressure environment, then we can ignore particle diffusion and assume that $\sigma_{surf}$ has a uniform value over the entire surface of the crystal, with $\sigma_{surf} > 0$ usually being provided by a far-away ice reservoir with $T_{reservoir} > T_{crystal}$.

Numerous experimental observations have shown that basal growth is generally slower than prism growth at low $\sigma_{surf}$ for temperatures above -3 C, owing to the large basal nucleation barrier that exists even at high temperatures close to 0 C [2013Lib, 2021Lib]. For this reason, we typically assume $dR_{thick}/dt \approx 0$ in the discussion that follows. Removing this assumption would not change our overall conclusions appreciably, as the interesting faceting behaviors are generally restricted to the prism facets at high temperatures.

To simplify our model further, we assume that the growth morphology is roughly stable in time, by which we mean that $r_{corner}$ does not change appreciably as $R_{tip}$ and $R_{facet}$ increase. This stability assumption cannot be absolutely accurate for a developing crystal over long periods of time, but approximately stable growth of this nature is frequently observed with slowly growing ice prisms, which is the primary focus of this paper.

Our assumption of stable growth allows us to write

$$\frac{dR_{tip}}{dt} \approx G_0 \frac{dR_{facet}}{dt} \qquad (16)$$

where $G_0$ is a geometrical constant that depends on the prism morphology. For a perfect hexagon, $G_0 = 2/\sqrt{3} = 1.155$, while $G_0 = 1$ for a circular shape. We have found that the exact value of $G_0$ between these limits has a negligible effect on our overall model predictions.

Ignoring crystal heating for the moment, Equation 5 becomes

$$v_{tip} \approx \alpha_{tip} v_{kin}(\sigma_{surf} - d_{sv}\kappa_{tip}) \qquad (17)$$

describing the tip growth with $v_{tip} = dR_{tip}/dt$, where typically $\alpha_{tip} \approx 1$ for the rounded tip surface. Likewise the prism facet growth is given by

$$v_{facet} \approx \alpha_{facet} v_{kin}(\sigma_{surf} - d_{sv}\kappa_{facet}) \qquad (18)$$

For the special case of a nearly isometric ice prism, we have $R_{thick} \approx R_{facet}$, giving

$$\kappa_{tip} \approx \frac{1}{R} + \frac{1}{r_{corner}} \qquad (19)$$

$$\kappa_{facet} \approx \frac{2}{R} \qquad (20)$$

where $R \approx R_{thick} \approx R_{facet}$ is the effective radius of the isometric prism. Note that $\kappa_{facet}$ refers to the center of a prism facet (bordered by two prism/prism edges and two prism/basal edges), while $\kappa_{tip}$ refers to the center of a prism/prism edge (midway between the two basal surfaces). With these assumptions, our stability condition in Equation 16 becomes

$$\alpha_{tip}(\sigma_{surf} - d_{sv}\kappa_{tip}) \\ \approx G_0 \alpha_{facet}(\sigma_{surf} - d_{sv}\kappa_{facet}) \qquad (21)$$

This expression is the primary result from our basic analytic model, and it shows that the degree of prism faceting is driven largely by the difference between $\kappa_{tip}$ and $\kappa_{facet}$ (for this special case of a nearly isometric hexagonal prism growing in a low-pressure environment). Because $\alpha_{tip} \approx 1 \gg \alpha_{facet}$ in most slow-growth conditions, the left side of Equation 21 would nearly always be larger than the right side if not for the difference between $\kappa_{tip}$ and $\kappa_{facet}$.

For example, starting with a perfectly sharp hexagonal prism, $r_{corner} \to 0$ and $\kappa_{tip} \to \infty$, producing a strong suppression of $v_{tip}$ via the Gibbs-Thomson effect, causing the prism corner to round. As $r_{corner}$ increases during this process, the Gibbs-Thomson effect lessens until the two sides of Equation 21 balance. Similarly, starting with a spherical crystal having $r_{corner} = R$ usually gives $v_{tip} > v_{facet}$, which causes the prism corner to sharpen. Regardless of the starting point, we see that Equation 21 describes the condition for stable growth with a value of $R/r_{corner}$ that is roughly independent of time.

Assuming a stable morphology can be achieved, Equation 21 can be rewritten to yield the model prediction

$$\frac{R}{r_{corner}} \approx \frac{R}{d_{sv}}\left(\sigma_{surf} - \frac{G_0 v_{facet}}{\alpha_{tip} v_{kin}}\right) - 1 \quad (22)$$

We recognize that the above approximations for $\kappa_{tip}$ and $\kappa_{facet}$ are only approximate, but we believe that they capture the underlying physics reasonably well. The values of these two parameters have the right order of magnitude, and their difference yields stable growth driven by sensible assumptions. Thus, while this basic analytic model is not absolutely accurate, we believe that it is useful for examining the physics underlying the faceting process over a broad range of conditions.

## Heat diffusion

When evaluating these equations in realistic experimental situations, one result that quickly appears is that latent heating must be incorporated into any model of the growth of simple ice prisms on a substrate in a near-vacuum environment [1972Lamb, 2021Lib]. Rather than resorting to full 3D finite-element diffusion calculations, we approximately incorporated latent-heating effects into our model by replacing Equations 17 and 18 with

$$v_{tip} \approx \alpha_{tip,tot} v_{kin}(\sigma_{surf} - d_{sv}\kappa_{tip}) \quad (23)$$

and

$$v_{facet} \approx \alpha_{facet,tot} v_{kin}(\sigma_{surf} - d_{sv}\kappa_{facet}) \quad (24)$$

where

$$\alpha_{tip,tot} = \frac{\alpha_{tip}\alpha_{the}}{\alpha_{tip} + \alpha_{therm}} \quad (25)$$

and

$$\alpha_{facet,tot} = \frac{\alpha_{facet}\alpha_{therm}}{\alpha_{facet} + \alpha_{therm}} \quad (26)$$

which then yields the modified stability condition

$$\frac{R}{r_{corner}} \approx \frac{R}{d_{sv}}\left(\sigma_{surf} - \frac{G_0 v_{facet}}{\alpha_{tip,tot} v_{kin}}\right) - 1 \quad (27)$$

Because $\alpha_{tip,tot} < \alpha_{tip}$ in most circumstances, this expression immediately shows that latent heating tends to decrease $R/r_{corner}$ at any given value of $v_{facet}$.

Once again, Equation 27 is not meant to be an exact expression, as we made several rather crude approximations regarding the Gibbs-Thomson effect and latent heating. Despite its shortcomings, however, we have found that this basic analytic model is quite useful for approximating the essential physics in various scenarios. Below we examine what this model says about ice crystal faceting as a function of temperature and other experimental parameters.

## Particle diffusion

We also considered the effects of particle diffusion on our model, using Equation 12 to make a rough estimate of the supersaturation field around a growing crystal. For a nearly isometric prism growing in conditions with $\alpha < \alpha_{diff}$, we found that the main difference between $\sigma_{surf}$ and $\sigma_\infty$ arose from the general trend in $\sigma(r)$ surrounding the crystal. Because an isometric hexagonal prism is not too different from a spherical shape, particle diffusion yielded only a modest difference between $\sigma_{surf,facet}$ and $\sigma_{surf,tip}$. Moreover, this difference was roughly proportional to pressure, while our main interest was comparing with crystal growth experiments done at low pressure. For these reasons, we found that the attachment kinetics and heat diffusion were the main drivers of faceting behavior (at low pressures), to the point that particle diffusion effects could be neglected without changing our main scientific conclusions.

## Thin plates in air

Our model is not much changed if we abandon the assumption of nearly isometric ice prisms, and it is especially useful to consider the growth of thin hexagonal ice plates in air, as such structures are a commonly observed in experiments in air at temperatures above -3 C.

For the thin-plate case with $R_{thick} \ll R_{facet}$, our surface-curvature terms become

$$\kappa_{tip} \approx \frac{1}{R_{thick}} + \frac{1}{r_{corner}} \quad (28)$$

$$\kappa_{facet} \approx \frac{1}{R_{thick}} + \frac{1}{R_{facet}} \quad (29)$$

while the stability condition in Equation 21 remains unchanged, yielding

$$\frac{R_{facet}}{r_{corne}} \approx \frac{R_{facet}}{d_{sv}} \left( \sigma_{surf} - \frac{G_0 v_{facet}}{\alpha_{tip} v_{kin}} \right) - \frac{R_{facet}}{R_{thick}} \quad (30)$$

which reduces to Equation 22 for an isometric prism.

For thin plates growing in air, however, we often have $d_{sv}/R_{facet} \ll \sigma_{surfeff}$ and $\alpha_{facet} \ll \alpha_{tip}$, where $\sigma_{surfeff} = (\sigma_{surf} - d_{sv}/R_{thick})$. And in these limits the stability condition becomes simply

$$r_{corner} \approx \frac{d_{sv}}{\sigma_{surfeff}} \quad (31)$$

independent of $R_{facet}$ and $R_{thick}$, while $v_{facet} \approx \alpha_{facet} v_{kin} \sigma_{surfeff}$.

## Higher-order effects from heat and particle diffusion

The above discussion focuses mainly on diffusion effects that arise from the lowest-order spatial changes in the temperature profile in a growing crystal (at low background gas pressure on a substrate) and in the supersaturation field around a growing crystal (at higher pressures, typically one atmosphere).

Extending this to higher-order effects, we find that heat diffusion tends to suppress corner growth, thus further reducing $R/r_{corner}$ relative to that calculated using the pseudo-1D model above. This follows because the corners of a nearly isometric faceted crystal growing on a substrate will experience the most latent heating on the crystal, as their thermal path to the substrate is the longest. The higher temperature rise at the corners suppresses their growth relative to other surfaces, thus increasing $r_{corner}$ and reducing $R/r_{corner}$.

In contrast, higher-order particle diffusion effects produce the opposite tendency, increasing $R/r_{corner}$ relative to that calculated using a 1D model. This comes about because of the Mullins-Sekerka instability [1964Mul], which tends to sharpen corners in particle-diffusion-limited growth.

The main takeaway from these paragraphs is that our basic model of faceting described above will likely overestimate $R/r_{corner}$ when significant heating is present, while underestimating $R/r_{corner}$ when particle diffusion is important. However, full 3D diffusion calculations are needed to fully quantify these statements in both cases.

## ❄ Model Predictions

We now examine some predictions from this analytic model by evaluating Equation 27 as a function of various experimental parameters. Beginning with the case of small isometric ice prisms growing in a near-vacuum environment, we choose the parameters:

$$R_{facet} = R_{thick} = R = 20 \text{ μm}$$
$$\alpha_{therm} = (10 \text{ μm})/R = 0.5$$
$$d_{sv} = 1 \text{ nm}$$
$$G_0 = 1.155$$

Our model evaluation begins by defining a table of $\sigma_{surf}$ values and calculating $v_{facet}$ using

$$\alpha_{facet} = A_1 e^{-\frac{\sigma_{0,1}}{(\sigma_{surf} - d_{sv}\kappa_{facet})}} + A_2 e^{-\frac{\sigma_{0,2}}{(\sigma_{surf} - d_{sv}\kappa_{facet})}} \quad (32)$$

with $\kappa_{facet} = 2/R$, giving

$$v_{facet} = \frac{\alpha_{facet}\alpha_{therm}}{\alpha_{facet} + \alpha_{ther}} \times$$
$$v_{kin}(\sigma_{surf} - d_{sv}\kappa_{facet}) \qquad (33)$$

Along with

$$\alpha_{tip,tot} = \frac{\alpha_{therm}}{1 + \alpha_{ther}} \qquad (34)$$

this gives us all we need to calculate $R/r_{corner}$ as a function of $v_{facet}$.

Table 1 shows the parameters we used for the prism attachment kinetics, which were chosen from experimental measurements of ice growth rates as a function of temperature and supersaturation [2021Lib]. We believe that these parameters are fairly accurate at the lowest temperatures but become more uncertain at the temperature increases. Using the sum of two nucleation processes is a convenient parameterization to include what we have called the "SDAK-2" phenomenon at the higher temperatures [2021Lib]. This phenomenon is speculative at present, and more work is needed to sort out the prism attachment kinetics at high temperatures. However, the parameters in Table 1 yield reasonable representations of the available measurements, and we believe that the remaining uncertainties in the details do not greatly affect the model results.

| T (C) | vkin (μm/sec) | A1 | sig0,1 | A2 | sig0,2 |
|---|---|---|---|---|---|
| -1 | 690 | 0.3 | 3e-5 | 0.7 | 1e-3 |
| -2 | 635 | 0.25 | 3e-4 | 0.75 | 1.5e-3 |
| -3 | 585 | 0.2 | 1e-3 | 0.8 | 3e-3 |
| -5 | 496 | 0.2 | 2e-3 | 0.8 | 5.5e-3 |
| -7 | 419 | 0.5 | 8e-3 | 0.5 | 1e-2 |
| -15 | 208 | 1 | 3e-2 | - | - |

Table 1. The parameters used to describe the prism attachment kinetics at different temperature using Equation 32.

Applying these various inputs yields the curves shown in Figures 2 and 3, which demonstrate the overall trends seen with our model. In both graphs we have calculated $R/r_{corner}$ as a function of $v_{facet}$ for several ice growth temperatures. The quantity $R/r_{corner}$ serves as a reasonable proxy for the overall degree of faceting, as $R/r_{corner} \rightarrow 1$ for an unfaceted (round) crystal exhibiting no prism faceting, and $R/r_{corner} \gg 1$ for a crystal exhibiting pronounced prism faceting. As mentioned above, strong basal faceting was assumed in our model from the outset, based on experimental observations.

Figure 2 shows first that the degree of prism faceting depends strongly on growth temperature, with more pronounced faceting at lower temperatures. This phenomenon clearly derives from the temperature dependence of the attachment kinetics on the prism facets, as measurements indicate that $\sigma_{0,prism}$ increases strongly with decreasing temperature, as indicated in Table 1 [2021Lib]. The overall trend in $\sigma_{0,prism}$ is well supported by experiments at temperatures below -2 C, while measurements at higher temperatures are more uncertain.

Figure 2 also shows that $R/r_{corner}$ increases with increasing $v_{facet}$, and this can be understood from the Gibbs-Thomson effect applied to the prism tips. The tip radius decreases as $\sigma_{surf}$ increases, resulting from a change in the balance between tip and facet growth needed to sustain a stable growth morphology, as described above.

We also see that latent heating is essentially negligible at the lowest growth rates, as one would expect, becoming important as the growth rate increases (for the assumed near-vacuum growth conditions). Latent heating is also generally more important at higher temperatures, which can be understood by examining how $\alpha_{therm}$ changes with temperature [2021Lib].

Finally, Figure 2 shows that prism faceting at low temperatures persists down to quite low growth rates, even though our model assumes a spherical ECS. This happens because a finite nucleation barrier yields a facet growth rate that decreases exponentially with the applied supersaturation, which does not happen on the rough tip surface. This phenomenon suggests that it will be difficult, from an experimental perspective, to "grow" an ECS in an environment with $\sigma_{surf} > 0$.

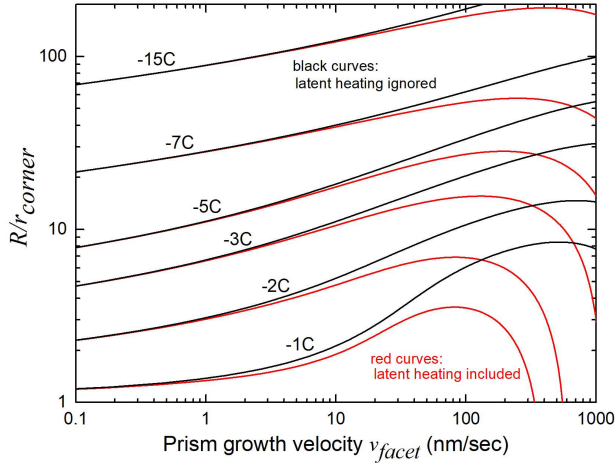

Figure 2. This graph shows the degree of prism faceting (as quantified by $R/r_{corner}$) as a function of the prism growth velocity, showing crystals with $R = 20$ μm growing at several different temperatures. The black curves ignore latent heating, while the red curves include this effect in the model. These curves show that: 1) prism faceting is most pronounced at lower temperatures, and 2) prism faceting persists even at quite low crystal growth rates, even though the model assumed a spherical ECS.

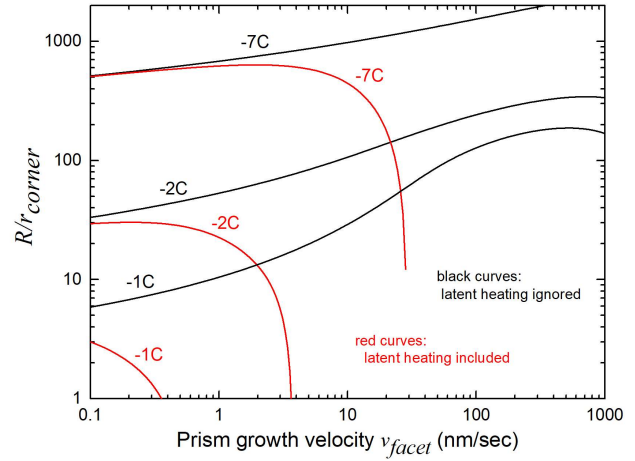

Figure 3. This shows the same model calculations as in Figure 2, but this time examining larger crystals with $R = 500$ μm. Here we see that faceting is more pronounced than in Figure 2 (higher $R/r_{corner}$ values) if latent heating is ignored, but the overall effects of latent heating are much greater with the larger crystals.

Moving our attention to Figure 3, we see that the values of $R/r_{corner}$ are higher with larger crystals, provided one ignores latent heating. This follows simply because $r_{corner}$ is roughly constant at a given growth rate, so the ratio $R/r_{corner}$ goes up with larger $R$.

Figure 3 also shows that the effects of latent heating are much more pronounced with larger crystals, as one would expect because $\alpha_{therm}$ decreases at $1/R$.

We adjusted the various parameters in our model to examine how this affects the plots shown in Figures 2 and 3. We found that the red curves were sensitive to our choice of $\alpha_{therm}$, meaning that our model of latent heating confirms our expectations that 1) thermal effects can be quite significant, and 2) our model only gives a rough estimate for how heating affects the degree of faceting. We also found that these results are only as good as our estimated parameters for the attachment kinetics, as one would expect. Even with these model uncertainties, however, the overall trends seen in Figures 2 and 3 appear to be quite robust with respect to modest changes in input parameters.

Note that we must have $R/r_{corner} \to 1$ as $v_{facet} \to 0$ in this model because we assumed a spherical ECS at all temperatures. This behavior is indeed seen, but remarkably low values of $v_{facet}$ must be obtained before $R/r_{corner} \to 1$ at low growth temperatures. This makes sense, as described above, because the nucleation barrier yields extremely low growth rates when $\sigma_{surf} \ll \sigma_0$.

## Time needed to realize the ECS and stable growth forms

It is instructive to examine how long it takes to achieve a stable growth morphology, as supplying the requisite time is not always practical in ice-growth experiments. A lower limit can be estimated from how long it takes a circular seed crystal to "fill out" into a faceted hexagonal prism, assuming that the final state has a large value of $R/r_{corner}$.

In this case the faceting time is mainly determined by the initial growth rate of the non-faceted corner before it becomes sharp, giving roughly

$$\tau_{facet} \approx \frac{G_1 R}{v_{kin}(\sigma_{surf} - d_{sv}\kappa_{facet})} \quad (35)$$

where $G_1$ is a geometrical factor of order unity and we assumed $\alpha_{tip} \gg \alpha_{facet}$ and $\alpha_{tip} \approx 1$. Because $\tau_{facet}$ is suitably small when considering the experimental observations described below, it is reasonable to expect that these ice crystals all had time to reach, or nearly reach, a stable growth morphology.

Note that the "fill out" time in Equation 35 is generally much shorter than the time needed to reach the ECS, which is approximately [2012Lib2]

$$\tau_{equilibrate} \approx \frac{R^2}{2\alpha v_{kin} d_{sv}} \quad (36)$$

For the case of a faceted ice prism relaxing to a spherical ECS, we see that $\tau_{equilibrate}$ is inversely proportional to $\alpha_{facet}$, which can result in extremely long equilibration times in practical ice-growth experiments.

## ❄ Numerical Modeling

While the stable-growth model described above is quite useful for examining faceting in slowly growing snow crystals over a broad range of conditions, it is not ideal for making detailed comparisons with targeted experimental investigations. If an experiment carefully defines the initial conditions, seed crystal morphology, growth conditions, and then measures the faceting behavior as a function of time, then a general dynamical model would provide a better way to compare experimental measurements with the more basic elements of our faceting model, including our initial assumption of a spherical ECS.

Creating such a dynamical model using the approximations and formalism described above is straightforward, as we simply need to evaluate the various growth velocities and then propagate the crystal forward in time, taking advantage of the relatively simple dynamics of the hexagonal prism morphology.

For prism geometry in Figure 1, the relation

$$R_{tip} = \frac{2}{\sqrt{3}} R_{facet} + \left(1 - \frac{2}{\sqrt{3}}\right) r_{corner} \quad (37)$$

ties the parameters together, giving $dr_{corner}/dt$ from the calculated $v_{tip}$ and $v_{facet}$. The model could easily be extended to include nonzero basal growth as well. With such a numerical model, one could drop the static-growth assumption we made above to examine a variety of time-dependent aspects of prism growth dynamics in detail.

Unfortunately, this model is limited by our approximate treatment of thermal and particle diffusion, which relied on analytic solutions to the spherical growth problem. Using a full 3D finite-element diffusion model is feasible for the relatively simple hexagonal-prism morphology [2001Woo], but do so is beyond the scope of this paper. Our main objective here is the generally simpler task of quantifying prism faceting behaviors over a broad range of growth conditions.

## ❄ Comparison with Ice-growth Experiments

Looking through our own ice-growth data archives, we have several prior experiments that have observed approximately stable simple prism growth over a range of conditions, which can be compared directly with the stable-growth model presented above. Moreover, there are earlier results in the literature that also lend themselves to possible reinterpretation using this model. As mentioned at the outset of this paper, our overarching goal is to develop a comprehensive picture of how facets develop in snow crystal growth as a function of temperature and other growth conditions, while better understanding the connections between slowly growing crystals and the equilibrium crystal shape.

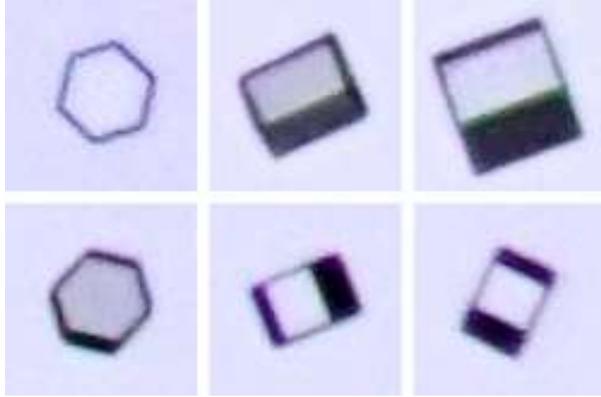

Figure 4. A representative sample of ice crystals growing at -7 C from water vapor in near vacuum conditions with an air pressure of 50 Torr. Each square image box is 50 μm on a side, and the crystals grew on a temperature-controlled sapphire substrate. Growth times were about 60 seconds with growth rates of about 150 nm/sec. Robust faceting appears under these conditions, exhibiting sharp edges and corners. The VPG apparatus used to grow these crystals is described in [2021Lib].

## Faceting below -5C

Numerous researchers have documented the formation of sharply faceted ice prisms in near-vacuum conditions at temperatures below -5 C [1982Gon, 1983Bec], and Figure 4 shows some representative examples. While the image resolution here is not sufficient to measure $R_{facet}/r_{corner}$ accurately, suffice it to say that these crystals exhibit a simple hexagonal-prism morphology with pronounced faceting and little rounding of the edges and corners. This growth behavior is easily explained from the model curves in Figure 2.

## Low-pressure observations with T ≥ -2C

At temperatures of -2 C and above, the model curves in Figure 2 predict smaller values of $R_{facet}/r_{corner}$ compared to crystals grown at lower temperatures, mainly resulting from the reduced nucleation barrier on prism facets at the higher temperatures. At -2 C, for example, Figure 5 shows clear prism faceting, as we expect from Figure 2, but now the edges and corners exhibit significant rounding brought about from the Gibbs-Thomson effect.

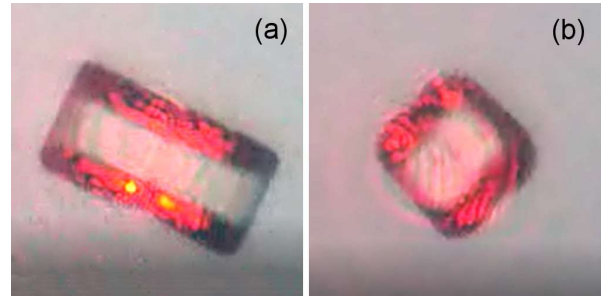

Figure 5. These images show a pair of ice crystals growing at -2C in a near-vacuum environment at 20 Torr on a sapphire substrate. (a) The overall size of this crystal is $(R, H) = (R_{facet}, R_{thick}) = (20$ μm, 37 μm) with a prism growth velocity of about 50 nm/sec. (b) This crystal has (R,H) = (19 μm, 18 μm) with a prism growth velocity of about 100 nm/sec. Both crystals grew from initially columnar seed crystals, and the red illumination is from a laser used to interferometrically measure the prism growth rates. Both crystals exhibit clear faceting at -2 C, but with some rounding of the corners. The VIG apparatus used to grow these crystals is described in [2021Lib].

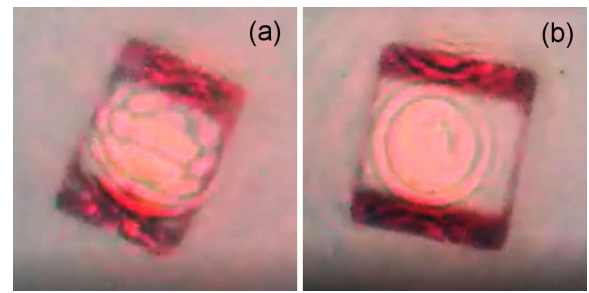

Figure 6. These images show a pair of example crystals growing at -1C in a near-vacuum environment on a sapphire substrate. The crystal in (a) has (R,H) = (29 μm, 21 μm) with a prism growth velocity of about 200 nm/sec, while (b) has (R,H) = (25 μm, 24 μm) with a prism growth velocity of about 100 nm/sec. Enhanced growth along the substrate in (b) yielded a form somewhat flatter than isometric, probably reducing the effects of latent heating. Both these crystals exhibit clear faceting at -1 C, but with some rounding of the corners. The VIG apparatus used to grow these crystals is described in [2021Lib].

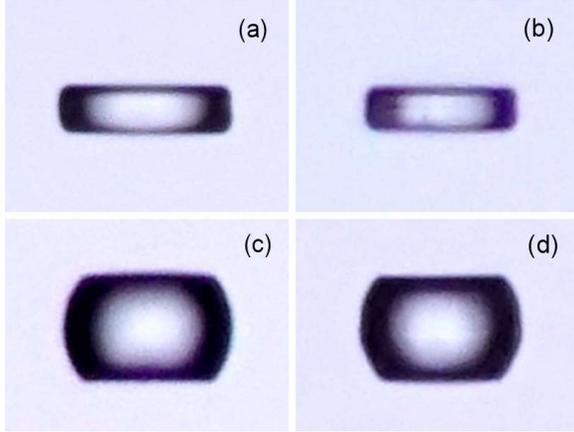

Figure 7. These images show additional examples of ice crystals growing at -2C in a near-vacuum environment on a sapphire substrate. Crystal (a) has (R,H) = (37 μm, 10 μm) with a prism growth velocity of about 100 nm/sec, while (b) has (R,H) = (23 μm, 7 μm) with a prism growth velocity of about 75 nm/sec. Rounding of the basal/prism edges obscures the prism facets somewhat in (a) and (b), because of the small H values. Crystal (c) has (R,H) = (28 μm, 18 μm) with a prism growth rate of 500 nm/sec, while (d) has (R,H) = (27 μm, 18 μm) with a prism growth rate of 650 nm/sec. At these faster growth rates, latent heating produces little faceting in (c) and (d), while basal faceting remains strong in all four crystals. The VPG apparatus used to grow these crystals is described in [2021Lib].

Figure 6 shows similar growth morphologies at -1 C, showing quite clearly that prism faceting can be quite prevalent at this temperature. The minor differences between the crystals in Figures 5 and 6 should not be taken too seriously at this point because the growth conditions and other parameters varied somewhat from crystal to crystal. As described in the figure captions, the crystals have different sizes and growth velocities, plus the initial conditions were not carefully noted at the time. We believe that these crystals provide representative examples of nearly stable growth forms, but better experiments are needed to fully document the growth and faceting behaviors as a function of time.

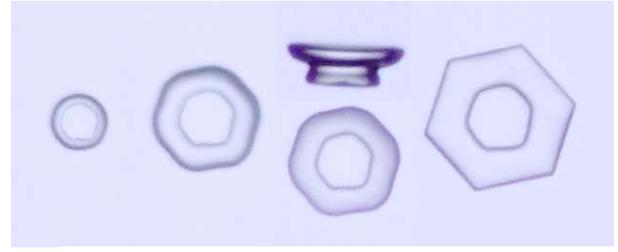

Figure 8. The four face-on crystals in the above composite image show four stages in the growth of a single plate-on-pedestal crystal in air at -2C and a pressure of one bar in the VPG apparatus [2021Lib]. The total growth time for this crystal was about 14 minutes, the effective radius of the final hexagonal plate was about 50 μm, and the final growth velocity was about 100 um/sec. The other crystal image in the composite shows a side view from the same set that appears to roughly correspond to the third image in the face-on set. The final face-on image shows a thin hexagonal plate with sharp prism facets growing out from a stout pedestal, with $R_{facet}/r_{corner}$ agreeing roughly with model predictions. The VPG apparatus used to grow these crystals is described in [2021Lib].

Figure 7 shows additional examples from a separate ice-growth experiment, and again we see that better targeted experiments will be needed to fully understand the subtle changes in faceting behaviors. Our model suggests stronger faceting for the crystals in Figures 7a and 7b, but it may be that rounding of the basal/prism edges may be obscuring the faceting somewhat in the images, owing to the smaller values of $R_{thick}$. In Figures 7c and 7d, our model suggests that the much higher growth rates for these crystals produced thermal effects that greatly diminished $R_{facet}/r_{corner}$ (as seen in Figure 2) and yielded crystal growth with essentially no observed prism faceting. In all cases, however, basal faceting is clearly seen, as expected.

## Growth in air with T ≥ -2C

Figure 8 shows a nice example of a thin plate growing in air at -2C, and the observed faceting in this crystal can be roughly explained by our model. In the first few images of the growth series, a small circular seed crystal initially develops a roughly circular plate-on-pedestal structure, and the plate appears to have rounded corners simply because it takes time for the corners to grow out as the plate edges becomes thinner.

After reaching its stable growth morphology, our model predicts $R_{facet}/r_{corner} \approx 20$, which is somewhat lower than seen in the final image. As described above, however, our model likely underestimates the value of $R_{facet}/r_{corner}$ when the growth is strongly limited by particle diffusion, as is certainly the case here showing growth in air at a pressure of one atmosphere.

Quantitative targeted experiments examining this kind of faceting transition in more detail could yield additional insights into the prism faceting process in air at these higher temperatures. Examining timeseries observations at different temperatures, supersaturations, and air pressures would likely improve our understanding of the attachment kinetics at high temperatures, provided computational growth models could adequately deal with the particle-diffusion problem.

Figure 9a presents another example illustrating that pronounced prism faceting can develop even at growth temperatures as high as -0.15 C. Our model cannot make detailed predictions of $R_{facet}/r_{corner}$ for this plate, as we have essentially no measurements of the attachment kinetics at such high temperatures. Turning this around, however, quantitative measurements like this could place interesting limits on $\alpha_{facet}$ for prism facets in this hard-to-observe growth regime.

Figure 9b illustrates another example of plate-on-needle growth in air at -1C, this time yielding a thicker plate. Observations like this reveal a remarkably rich diversity of growth morphologies as the temperature and supersaturation are varied [2021Lib2]. The biggest challenge in interpreting the observations lies in producing computational models that are capable of accurately handling diffusion-limited growth in the presence of highly anisotropic attachment kinetics [2021Lib].

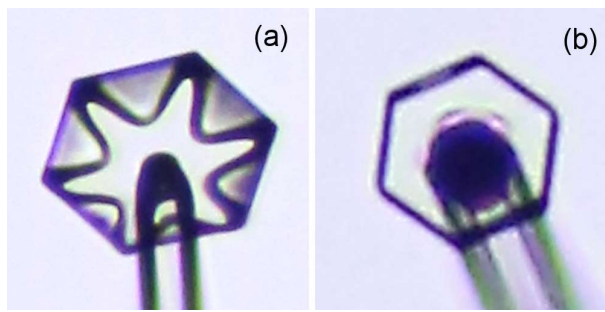

Figure 9. These two images show ice plates growing on the ends of slender ice needles in air. The crystal in (a) was grown at -0.15 C within a background supersaturation of 8%, the effective plate radius was about 55 μm, and the prism growth velocity was about 85 nm/sec. The thin platelike hexagonal morphology clearly demonstrates that prism faceting can develop even at temperatures as high as -0.15 C. The crystal in (b) was grown at -1 C with a background supersaturation of 8%. The effective plate radius was about 35 μm, and the prism growth velocity was about 70 nm/sec. The Dual-Chamber apparatus used to grow these crystals is described in [2021Lib].

In contrast to the thin snow-crystal plates growing from water vapor seen in Figures 8 and 9, Figure 10 shows a nice example of a circular disk growing from slightly supercooled liquid water. This well-known phenomenon [2003Shi, 2005Shi] indicates that the prism attachment kinetics at the ice/water interface is highly isotropic. In contrast, the prism attachment kinetics at the ice/vapor interface clearly retains some anisotropy as one approaches 0C.

Basal facets are seen in both the ice/water and ice/vapor systems, suggesting a correspondence between the nonzero step energies at both interfaces [2014Lib]. The different behaviors of prism surfaces at the ice/water and ice/vapor systems require some other explanation. Adding to the mystery, the step energies on prism facets tend toward zero at 0 C for both ice/vapor and ice/liquid. We have postulated a "frustrated" growth model that might explain the observations [2021Lib], but additional work will be needed to fully

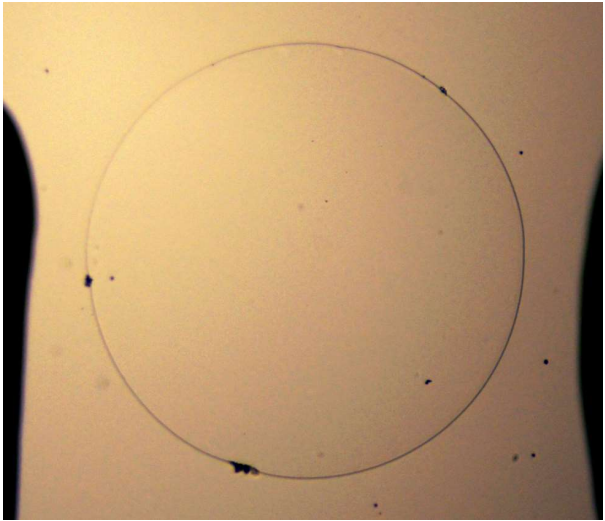

Figure 10. This photo shows a 2-mm-diameter disk of ice growing outward on the surface of a thin film of slightly supercooled water covering a glass plate. The c-axis of the oriented ice crystal is aligned perpendicular to the glass surface. The large dark regions are copper support arms glued to the glass, while dark specks are dust particles in the water film.

explain prism faceting in these two systems near the triple point.

### Surface roughening?

The investigation described here was substantially motivated by Elbaum's paper [1991Elb] describing a surface roughening transition on the prism facet of ice, so we next examine this result in some detail. The crystals described in that paper were quite large, growing in near vacuum conditions, so we consider a specific example of an isometric crystal with $R_{facet} = 500$ μm and $v_{facet} = 5$ nm/sec, as these parameters approximately correspond to the primary example described in Figure 3 in [1991Elb].

Applying our stable-growth model to crystals of this size gives the results in Figure 3 above, where we see that latent heating is expected to have a large effect on prism faceting. For crystals growing at a fixed velocity of 5 nm/sec, our model predicts a rather abrupt transition in faceting behavior at a temperature around -2 C. From the red curves in Figure 3, we see that $R_{facet}/r_{corner}$ transitions from large values at -7 C to $R_{facet}/r_{corner} \approx 1$ at temperatures above -2 C. In contrast to [1991Elb], however, we find that this transitional behavior can be explained by latent heating along with the surface attachment kinetics on prism surfaces, as these factors both change substantially with temperature.

Carrying this further, we suggest that latent heating may also explain the "domed" structure of the prism facets described in [1991Elb]. This slight deviation in flatness could arise from the same thermal gradients causing the overall rounding of the crystal morphology, although modeling such a result would require more information about the growing crystals and their environment. Our main conclusion here is that latent heating is an important factor affecting growth and faceting, and this factor was not carefully considered in [1991Elb].

Note also that the nearly flat (domed) facets described in [1991Elb] present a dynamical quandary at a more fundamental level. Because the facet surface was growing upward uniformly at 5 nm/sec, we must have $v_{facet} = v_{vicinal}$, where $v_{facet}$ describes the growth of the top prism terrace and $v_{vicinal}$ describes the surrounding vicinal surfaces. Applying Equation 1 then yields $\alpha_{facet}\sigma_{facet} = \alpha_{vicinal}\sigma_{vicinal}$, and any sensible model of the attachment kinetics gives $\alpha_{facet} < \alpha_{vicinal}$, thus giving $\sigma_{facet} > \sigma_{vicinal}$. This latter inequality is hard to avoid in any dynamical analysis of a growing "domed" surface, but it is easily explained by latent heating effects (which heat the corners more than the facet centers).

All these considerations cast doubt on Elbaum's scientific conclusion of a surface roughening transition. Our dynamical growth model provides a natural explanation of the observations, with reasonable model inputs, even while assuming an isotropic surface energy.

Looking at the bigger picture, we note that our growth model incorporates a decreasing step energy on the prism facets with increasing temperature, which could be interpreted as a gradual roughening transition (because a rough surface is equivalent to a surface with vanishing step energy). There is an important distinction to be made, however, in that our model assumes from the outset that the ECS is spherical at all temperatures. Put another way, our model assumes that the surface energy anisotropy is negligible, so it cannot be responsible for producing faceted forms. Instead, the changing step energy affects the *dynamics* of crystal growth via terrace nucleation, and this brings about faceted growth forms. A roughening transition usually refers to the *equilibrium* structure of the crystal surface.

## Observing the equilibrium crystal shape

Part of this discussion must deal with the problem of how difficult it is to observe the ice ECS in practical experiments. Based on measurements of the terrace step energies on basal and prism surfaces as a function of temperature [2013Lib, 2021Lib], we have argued that the available evidence suggests that the ice ECS is nearly spherical at temperatures above -15 C [2012Lib2, 2021Lib]. If one therefore assumes that the ECS is spherical, it quickly becomes apparent that observing this morphology in equilibrium is a challenging experimental task.

If one begins with a faceted growth form, then relaxing to the ECS would require that ice sublimate from the faceted corners and deposit on the facet surfaces until the spherical ECS is obtained. This process is greatly suppressed, however, by the extremely slow attachment kinetics on faceted surfaces at low supersaturations, as modeled in Equation 3. As quantified in Equation 36, the time needed to complete this equilibration to the ECS can be far longer than any experiment performed to date.

In both [1985Col] and [1991Elb], the authors described measurements of the ice ECS based on slowly growing crystals, assuming that the experimental wait times were sufficient to achieve the ECS. Our new model suggests, however, that achieving the ECS using slowly growing ice crystals may be nearly impossible if the true ECS is spherical. Referring to Figures 2 and 3, we see that growth forms remain faceted even at extremely low growth velocities, simply because $\alpha_{facet}$ goes to zero rapidly when $\sigma_{surf} \ll \sigma_0$. Given the experimental uncertainties in [1985Col, 1991Elb], we believe that the observations could easily be explained from our dynamical model with a spherical ECS. Moreover, we feel that no experiment to date has definitively observed the ice ECS.

## An ECS instability

Even at a fundamental theoretical level, it would not have been possible to observe the true ice ECS in any experiment performed to date. In all prior experiments, test crystals were grown in an environment with some $\sigma_\infty$ specified as a far-away boundary conditions, and no ECS can stably exist in such conditions.

To see this, consider a spherical crystal with some radius $R$ within such a growth chamber. The crystal would be in equilibrium (neither growing nor sublimating) provided $\sigma_\infty = d_{sv}\kappa$, as indicated in Equation 5. But this equilibrium state is not a stable state. If one perturbs the crystal to slightly increase $R$, then the equilibrium condition would not be met, and the crystal would begin growing. And it would continue growing indefinitely thereafter. Alternatively, perturbing the crystal to slightly decrease $R$ would cause sublimation that would continue until the crystal sublimated away completely.

What this shows is that no ECS can stably exist when a fixed outer boundary of $\sigma_\infty$ is maintained. The only way to produce a truly stable ECS is to isolate a single crystal in an otherwise empty environment, as then the background supersaturation will adjust to come into equilibrium with the ECS.

Reflecting on this discussion suggests that creating an isolated void in a single-crystal ice block would likely be the best approach to observing the ice ECS in the lab. A vacuum pump attached to a capillary needle could create a small void, and an applied temperature gradient could be used to move the void away from the capillary tip. Once isolated, a uniform temperature environment could be applied to allow the crystal to reach the ECS.

If the ECS were spherical, or nearly so, then an initially faceted void (the growth form of the void) [1965Kni, 1993Fur] would quickly evolve toward the ECS, as this evolution would not be hindered by any nucleation barriers. Moreover, applying a quadrupolar temperature profile would distort the shape of the void, thus allowing a measurement of the ice surface energy as a function of temperature. Realizing such an experiment is a task left for another day, but clearly there is substantial opportunity for improving our understanding of the ice surface energy, surface energy anisotropy, and the ice ECS.

## ❄ Conclusions

In summary, we have developed comprehensive dynamical model describing the growth of faceted prisms with rounded edges and corners. Our input model assumptions were guided by recent ice-growth measurements, including: 1) we assumed an isotropic surface energy and therefore a spherical ECS, 2) we assumed strong basal faceting and negligible basal growth rates for slowly growing crystals in a near-vacuum environment, and 3) we assumed prism faceting governed by a terrace-nucleation model, with nucleation parameters derived from growth measurements.

Our model uses approximate calculations for particle and heat diffusion to yield analytic expressions for growth morphologies in a stable-growth limit, as this approach allows reasonable estimates of faceting behaviors over a broad range of growth conditions. Dropping the stable-growth assumption, numerical modeling could be used to examine time-dependent morphological changes for comparison with targeted ice-growth experiments. A full 3D computation model describing particle and heat diffusion in the presence of strongly anisotropic attachment kinetics remains a challenging problem, not addressed in this paper.

Our scientific conclusions based on model calculations include:

- For ice crystals grown on a substrate in a near-vacuum environment, our model shows that latent heat diffusion can strongly affect growth rates and faceting behavior. These effects are especially strong with large crystals, at high temperature, and at high growth rates, as shown in Figures 2 and 3.

- Our relatively simple analytic model likely overestimates the value of $R_{facet}/r_{corner}$ when heat diffusion plays a major role, while it underestimates the value of $R_{facet}/r_{corner}$ when particle diffusion limits growth. Heat diffusion (for a faceted prism growing on a substrate in a near-vacuum environment) tends to result in the highest crystal temperatures at positions farthest from the substrate, yielding rounded corners and lower values of $R_{facet}/r_{corner}$. Particle diffusion tends to sharpen corners via the Mullins-Sekerka instability [1964Mul], thus yielding higher $R_{facet}/r_{corner}$ values. Incorporating these higher-order diffusion effects would require full 3D diffusion modeling.

- For large prisms ($R_{facet} \approx 500$ μm) growing at roughly 1-10 nm/sec, our model predicts an abrupt transition from sharply faceted prisms ($R_{facet}/r_{corner} \gg 1$) at temperatures below about -2 C to rounded forms ($R_{facet}/r_{corner} \approx 1$) at higher temperatures. Elbaum interpreted this faceting behavior as a roughening transition of the prism surface near -2 C [1991Elb], but we believe that our dynamical model provides a better explanation. In our picture, there is no roughening transition, and the ice ECS is essentially spherical at all temperatures above -15 C.

- Our model indicates that strong faceting (defined by large $R_{facet}/r_{corner}$ values) persists down to remarkably low growth rates, especially at low temps, as seen in Figures 2 and 3. This result suggests that the faceting behaviors described in [1985Col] could be explained reasonably well as a dynamical growth phenomenon. The result also suggests that it can be exceedingly difficult to observe the ECS using growing crystals, casting doubt on the conclusions described in [1985Col].

- Our model suggests that a strong anisotropy in the ice surface energy is not required to explain observations of faceted ice growth. In nearly all cases, the formation of ice-crystal facets appears to result from the strong anisotropy in the surface attachment kinetics.

- Using data from different ice-growth experiments, we find that all our existing observations of simple faceted forms are generally consistent with the growth model described above, which incorporates the comprehensive basal and prism attachment kinetics model described in [2021Lib]. From this we continue to build a self-consistent picture of the attachment kinetics and of snow crystal growth that can reasonably explain the most reliable experimental data. This evolving paradigm also serves to suggest targeted experimental investigations that can further influence and refine our broader understanding of the structure and molecular dynamics of the ice surface.

- There is much potential for making additional progress in understanding the dynamics of ice crystal growth using precision experiments measuring ice growth rates and morphological behaviors in different environments. Unfortunately, such investigations are substantially hampered at present by the lack of adequate computational techniques that can model crystal growth in the presence of strongly anisotropic attachment kinetics in combination with particle and/or latent-heat diffusion. As these computational tools become available, they will enable much improved comparisons between theory and experiment that will undoubtedly yield further insights into the physical processes underlying ice crystal growth dynamics.

We gratefully acknowledge support from the Cambridge-Caltech Exchange Program and the Summer Undergraduate Research Fellowship program at Caltech.

## ❄ References